# Estimation of presampling modulation transfer function in synchrotron radiation microtomography

Ryuta Mizutani [a,] *, Keisuke Taguchi [a], Akihisa Takeuchi [b], Kentaro Uesugi [b], Yoshio Suzuki [b]

[a] Department of Applied Biochemistry, School of Engineering, Tokai University, Kitakaname 4-1-1, Hiratsuka, Kanagawa 259-1292, Japan

[b] Research and Utilization Division, JASRI/SPring-8, Kouto 1-1-1, Sayo, Hyogo 679-5198, Japan

*Corresponding author. Tel: +81-463-58-1211; fax: +81-463-50-2506.
E-mail address: ryuta@tokai-u.jp (R. Mizutani).

**Abstract**

The spatial resolution achieved by recent synchrotron radiation microtomographs should be estimated from the modulation transfer function (MTF) on the micrometer scale. Step response functions of a synchrotron radiation microtomograph were determined by the slanted edge method by using high-precision surfaces of diamond crystal and ion-milled aluminum wire. Tilted reconstruction was introduced to enable any edge to be used as the slanted edge by defining the reconstruction pixel matrix in an arbitrary orientation. MTFs were estimated from the step response functions of the slanted edges. The obtained MTFs coincided with MTF values estimated from square-wave patterns milled on the aluminum surface. Although x-ray refraction influences should be taken into account to evaluate MTFs, any flat surfaces with nanometer roughness can be used to determine the spatial resolutions of microtomographs.

**1. Introduction**

The application of synchrotron radiation to microtomography has allowed high-resolution three-dimensional analysis [1-4]. High-resolution microtomography has been used to visualize a wide variety of three-dimensional microstructures ranging from crack propagation in metals [5] to neuronal networks of human brain [6]. The modulation transfer function (MTF) on the micrometer scale facilitates the estimation of the spatial resolutions of the microtomographs. We have determined the microtomographic MTF from square-wave patterns fabricated by focused ion-beam (FIB) milling [7,8]. The test patterns revealed spatial resolutions of the synchrotron radiation microtomographs, though nanometer precision patterns should be prepared to estimate spatial resolutions of the microtomographs.

The step response function is widely used for estimating MTFs. For computed tomography (CT), the step response function of a clinical CT has been measured using a flat resin surface slanted with respect to the reconstruction matrix [9]. A presampling profile of the step response of the slanted edge can be determined by plotting pixel densities versus the distance from the edge (Figure 1). The MTF is then obtained from the Fourier transformation of the line spread function (LSF) of the step response. Although the MTFs of clinical CTs can be estimated from test objects on the millimeter scale, the spatial resolutions achieved by recent synchrotron radiation microtomographs should be estimated by using high-precision edges with roughness much less than their spatial resolutions.

A crystal or ion-milled surface can be used as an edge for microtomographic MTF estimation. In this study, a flat surface of diamond crystal and an ion-milled edge prepared on an aluminum wire were used as slanted edges for MTF estimation. Square-wave patterns were also fabricated on these surfaces by FIB milling as references for the resolution estimation. Tilted reconstruction was introduced to enable any edge to be used as the slanted edge by defining the reconstruction pixel matrix in an arbitrary orientation. The resolution estimation from the diamond and aluminum test objects is discussed on the basis of the obtained MTFs.

## 2. Materials and Methods

### 2.1. Test objects

Most microtomographic analyses are performed under atmospheric conditions. Aluminum and diamond were chosen as test materials on the basis of their chemical stability in the atmosphere.

An aluminum test object for resolution estimation was fabricated by using an FIB apparatus (FB-2000; Hitachi High-Technologies, Japan) operated at 30 kV. An aluminum wire with a diameter of 250 μm was attached to the sample holder and subjected to gallium ion beam milling. A gallium beam of 4.5 nA was used for rough abrasion of a 50-μm wide region of the surface and a 1.7 nA beam was used to finish the flat surface. Although fabrication precision of approximately 50 nm was achieved with this method, striations with depths of 50–100 nm were observed on the surface, as shown in Figure 2a. Therefore, the surface roughness was estimated to be 100 nm. A series of square wells was also carved on this surface, as reported previously [8]. The pitches of the square-wave patterns were 2.0, 1.6, 1.2, 1.0, 0.8 and 0.6 μm. Each pattern was composed of half-pitch wells and half-pitch intervals, *i.e.*, 0.5-μm well and 0.5-μm interval for a 1.0-μm pitch.

A diamond crystal (IMS, size 100/120; Tomei Diamond, Japan) with approximate dimensions of 150 μm was attached to an aluminum wire tip with epoxy glue. The diamond crystal and the aluminum tip were bridged with an adhesive carbon tape to prevent the accumulation of electrostatic charge from the ion beam. Although small defects were observed on crystal surfaces, the surface roughness was estimated to be less than 50 nm from the secondary electron image shown in Figure 2b. A series of square wells was also carved on this surface. The pitches of the square-wave patterns were 2.0, 1.6, 1.2, 1.0, 0.8 and 0.6 μm. A gallium beam of 2.05 nA was used to carve the 2.0, 1.6, and 1.2 μm patterns and a 0.47 nA beam was used for the 1.0, 0.8 and 0.6 μm patterns, giving rectangular wells with depths of 3–5 μm. A secondary electron image of the pattern is shown in Figure 2b. The pattern structures were slightly affected by the charge accumulation from the ion beam.

These test objects were then recovered and mounted on stainless steel pins by using epoxy glue. The carbon tape attached to the diamond crystal was removed. Air exposure had no effect on the pattern structures.

*2.2. Microtomographic measurement*

The microtomographic analysis was performed at the BL20XU beamline [4,10] of SPring-8. The test object was mounted on the goniometer head of the microtomograph by using a brass fitting designed for the pin-hold sample. The eccentricity of sample rotation was estimated to be 0.2 μm. Transmission radiographs were recorded with a CCD-based x-ray imaging detector (AA50 and C4880-41S; Hamamatsu Photonics, Japan) using 12 keV x-rays. The distance between the fluorescence screen of the detector and the sample rotation axis was 7 mm. Each radiograph was taken by averaging pixels in 2 × 2 bins. The number of detector pixels after binning was 2000 in the horizontal direction perpendicular to the sample rotation axis and 1312 along the vertical axis. These radiographs with an effective pixel size of 0.50 μm × 0.50 μm were acquired in a parallel-beam geometry, giving the same pixel size at the sample position. Hence, the field of view was 1000 μm × 656 μm. The images of this area were acquired with a rotation step of 0.10° and exposure time of 300 ms per image.

Since the detector was placed in the proximity of the sample, x-ray reflections from the diamond crystal were also observed in the transmission image. The reflection gave bright regions in the sample images at certain crystal orientations, which will cause artifacts in microtomograms. In such circumstances, the detector-to-sample distance was slightly modified so that the reflection did not overlap with the sample image.

*2.3. Reconstruction calculation*

The microtomographic reconstruction was performed by using the Donner algorithm [11]. The convolution back-projection method with a Hann-window filter [11,12] was used for the reconstruction calculation. A double zoomed reconstruction was chosen to retain the high-resolution information [7].

The step response function was determined using flat surfaces of the test objects. The flat surfaces should be slightly slanted with respect to the reconstruction pixel matrix to obtain the presampling step response [9]. However, it is difficult to place micro test objects with an angular precision of a few degrees with respect to the reconstruction matrix. In the tomographic reconstruction, the axes of the reconstruction matrix were oriented along the direction from which the tomographic scan started. These axes are not defined by principle but chosen artificially. Therefore, the axes can be arbitrarily tilted within the tomographic slice plane. Assuming that the reconstruction matrix is tilted at angle $\alpha$, the linear absorption coefficient $\mu(x, y)$ can be calculated as an integral with respect to the scan angle $\theta$:

$$\mu(x,y) = \int_0^\pi p'(x\cos(\theta+\alpha) + y\sin(\theta+\alpha), \theta)d\theta, \qquad (1)$$

where $p'(\omega, \theta)$ indicates the inverse Fourier transform of the filtered projection strip at angle $\theta$ and lateral position $\omega$. This corresponds to a rotational scan starting from angle $\alpha$. Even after data acquisition, the reconstruction matrix can be oriented at any angle to the edge surface by setting tilt angle $\alpha$. In this study, with this tilted reconstruction method the edge surfaces were slanted by approximately 5° from the reconstruction matrix.

These calculations were performed by using the program RecView (available from http://www.el.u-tokai.ac.jp/ryuta/) accelerated with CUDA parallel-computing processors.

*2.4. MTF estimation*

The MTF estimation from the slanted edge was performed as reported previously [9]. The ion-milled surface of the aluminum test object was placed at an inclination of 6.5° to the column line of the reconstruction matrix, and the diamond surface was placed at 5.6°. The in-plane step responses were calculated using 360 voxels. The MTF was estimated from the Fourier transformation of the LSF of the step response.

The MTF values were also estimated from the square-wave patterns carved along the in-plane direction. The estimation was repeated for five tomographic slices of the square-wave pattern to determine the deviation in the MTF value. These statistical calculations were performed using Microsoft Excel.

## 3. Results and Discussion

### 3.1. MTF estimation

The step response functions of the diamond and aluminum test objects are shown in Figure 3. The LSFs determined from the step response are also superposed. The full widths at half maximum (FWHM) of the LSFs of diamond and aluminum were both estimated to be 0.8 μm. Although the upper and lower ends of the step response should converge to flat profiles, x-ray refractions observed at the edge surface distorted the step ends. The resultant LSF consisted of a central peak derived from the x-ray absorption and lateral undershoots caused by the x-ray refraction. The lateral undershoots in the diamond LSF appeared large since the linear absorption coefficient (LAC) of diamond is comparatively low (4.6 $cm^{-1}$) at 12 keV. The aluminum LSF also suffered from refractions, but the lateral undershoots were small in comparison with the central peak. This is because the aluminum LAC is nine times higher (41.3 $cm^{-1}$) than that of diamond, while the real parts of the deviation of the refractive index from unity are similar in aluminum and diamond.

The edge enhancement in the tomogram "with not too high spatial resolution" is represented by three-dimensional Laplacians of the real part of the refractive index [13]. However, the edge enhancement in the high-resolution microtomogram of this study could not be represented simply by the Laplacians of the refractive index. Therefore, step responses were simulated by using the Kirchhoff integral, as described in the Appendix [14-16]. The obtained aluminum response is also superposed in Figure 3. Although the step responses were simulated without the tomographic reconstruction process, the simulated aluminum response agreed with the observed response. This indicates that the edge enhancement in the microtomogram of the aluminum test object quantitatively represents the refraction effect. In contrast, large edge fringes simulated in the diamond step response were not observed in the microtomogram. Since the refraction effect predominant in the diamond response depends on the edge shape of the test object, the discrepancy between the observed and simulated responses can be ascribed to the crystal surface roughness.

The MTFs estimated from the aluminum and diamond LSFs are shown in Figure 4. The high-frequency MTFs of the aluminum and diamond test objects gave a spatial resolution of 0.9 μm at 5% of the MTF. This is comparable to the resolution of two-dimensional images, which was estimated to be 1 μm [17]. It has been discussed that the image resolution depends on the thickness of the fluorescence screen of the imaging detector [4]. The low-frequency response of the MTF gave values above unity mainly because of the refraction effect. This overestimation was larger in the diamond MTF since the lateral undershoots caused by the refraction were comparatively large in the diamond LSF.

Microtomograms of the square-wave patterns are shown in Figure 5. The patterns with a pitch of up to 1.0 μm were resolved for both the aluminum and diamond test objects. The variance and amplitude of the aluminum pattern were used for calculating the MTF values [18,19]. The obtained MTF values coincide with the MTF from the step response, as shown in Figure 4, indicating that the slanted edge method can be applied to the MTF estimation of microtomographs.

*3.2. Slanted edge method in microtomography*

The secondary electron image of the diamond crystal indicated that the crystal surfaces can be regarded as flat edges with nanometer roughness. Therefore, the MTF of microtomographs can be estimated by using crystal samples. The low LAC of diamond resulted in distortions of the step response function due to the x-ray refraction. The edge distortion was relatively small in the step response of the aluminum surface with the larger LAC step. Crystals with a sufficient LAC value would facilitate the resolution estimation of microtomographs from the step response of crystal surfaces.

MTFs can also be estimated from the step response of artificially prepared surfaces. Microtomographic resolutions should be estimated by using high-precision edges with roughness much less than the spatial resolution. Therefore, it is important to finish the surface with nanometer roughness. This can be achieved by using a low-current beam for the FIB apparatus. A high-precision edge might be prepared by cutting metal objects with sharp razor blades, though the surface roughness should be examined with an electron microscope. The LAC of the test object material should also be taken into account.

The results obtained in this study indicate that the slanted edge method [9] can be applied to the resolution estimation of synchrotron radiation microtomographs. Three-dimensional x-ray visualization by a holotomographic approach [20] and coherent diffraction imaging [21] have also been reported. MTFs of any three-dimensional visualization systems can be estimated with the slanted edge method by taking images of a test object with sufficient surface roughness. Edges with nanometer roughness can be found in samples subjected to three-dimensional analysis. In such cases, MTFs can be estimated using the slanted edge method by calculating the sample image with the tilted reconstruction. This approach allows *in-situ* estimation of MTFs by using any high-precision edges in the samples.

**Acknowledgements**

We thank Yasuo Miyamoto, Technical Service Coordination Office, Tokai University, for helpful assistance with FIB milling. This study was supported in part by Grants-in-Aid for Scientific Research from the Japan Society for the Promotion of Science (no. 21611009). The synchrotron radiation experiments were performed at SPring-8 with the approval of the Japan Synchrotron Radiation Research Institute (JASRI) (proposal nos. 2007A1844, 2007B1102, 2008A1190, 2009A1113 and 2009B1191).

**Appendix**

When an object is illuminated by a plane wave, the electric field $E(x)$ just behind the object can be written as

$$E(x) = \exp[-\beta(x) + i\Phi(x)], \tag{A.1}$$

where $x$ represents coordinate just behind the object, $\beta(x)$ is the absorption, and $\Phi(x)$ is the phase shift. The electric field at the image plane $F(y, r)$ is expressed by using the Kirchhoff integral [14] as

$$F(y,r) = \frac{1}{\sqrt{\lambda r}} \exp(-ikr) \int E(x) \exp\left[-\frac{ik(y-x)^2}{2r}\right] dx, \tag{A.2}$$

where $y$ represents the coordinate at the image plane, $r$ is the object-to-detector distance, and $k$ is a wave vector defined by $k=2\pi/\lambda$. The intensity profile at the image plane $I(y) = |F(y,r)|^2$ can be obtained by

$$I(y) = \frac{1}{\lambda r} \left| \int \exp\left[-\beta(x) + i\Phi(x) - \frac{ik(y-x)^2}{2r}\right] dx \right|^2, \tag{A.3}$$

where $I(y)$ is normalized to the beam intensity. The intensity profile in the detector image $I_d(y)$ should be calculated by the convolution of $I(y)$ and the observation LSF. Assuming that the observation LSF is approximated with a Gaussian function, the profile $I_d(y)$ is given by

$$I_d(y) = \frac{1}{\sqrt{\pi} s} \int I(y-u) \exp\left[-\left(\frac{u}{s}\right)^2\right] du, \tag{A.4}$$

where $s$ is proportional to FWHM of the LSF. The simulated step responses (Figure A1) were calculated with $s = 0.35$ μm, corresponding to an FWHM of 0.58 μm.

**Figure captions**

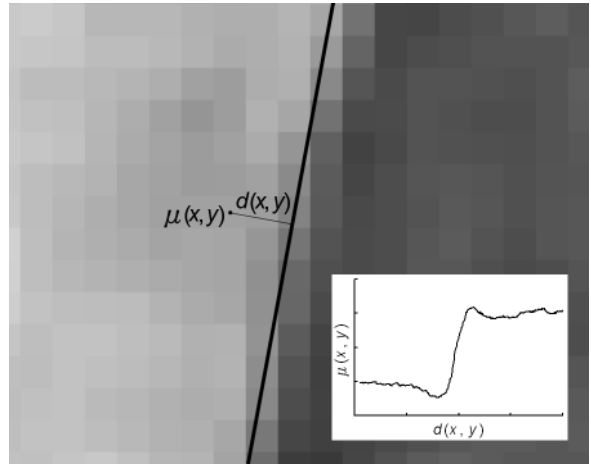

Figure 1. Schematic representation of a surface edge slanted with respect to the reconstruction matrix. The sample surface determined from the tomographic image is indicated with a thick line. Density $\mu$ of each pixel is plotted with respect to distance $d$ from the sample surface, giving the presampling step response shown in the inset.

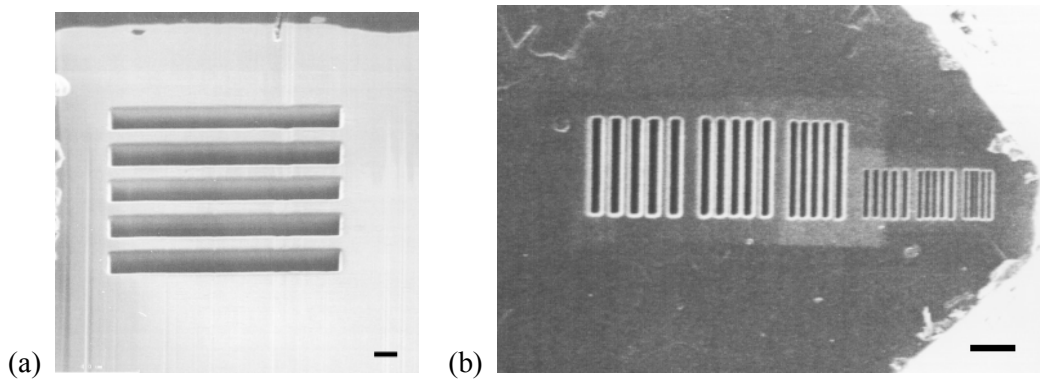

Figure 2. FIB secondary electron images of test objects. (a) A square-wave pattern with a 2.0 μm pitch carved on an aluminum surface. Scale bar: 1 μm. (b) Diamond test object. Square-wave patterns with pitches of 2.0, 1.6, 1.2, 1.0, 0.8 and 0.6 μm were carved on the crystal surface. Scale bar: 5 μm.

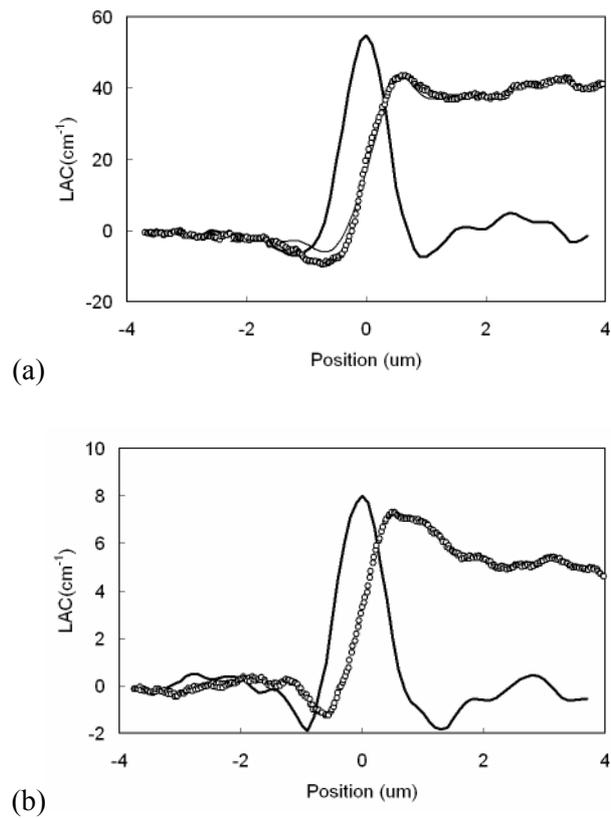

Figure 3. Step response functions and LSFs of aluminum (a) and diamond (b) test objects. The step responses are plotted with open circles. LSFs deduced from the step responses are drawn in thick lines. The simulated step response of 250 μm aluminum is drawn with a thin line.

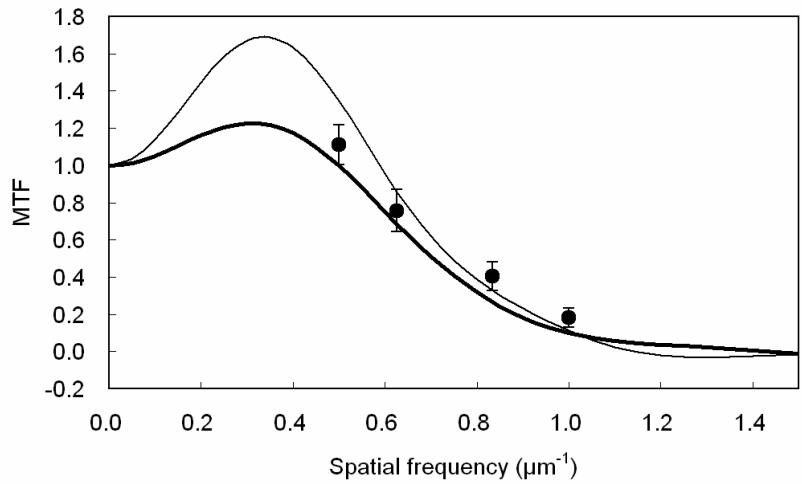

Figure 4. MTFs estimated from slanted edges and square-wave patterns. The thick line represents the MTF estimated from the LSF of aluminum surface and the thin line represents that from the diamond crystal surface. Closed circles represent MTF values estimated from square-wave patterns of the aluminum test object. Error bars indicate 95% confidence limits.

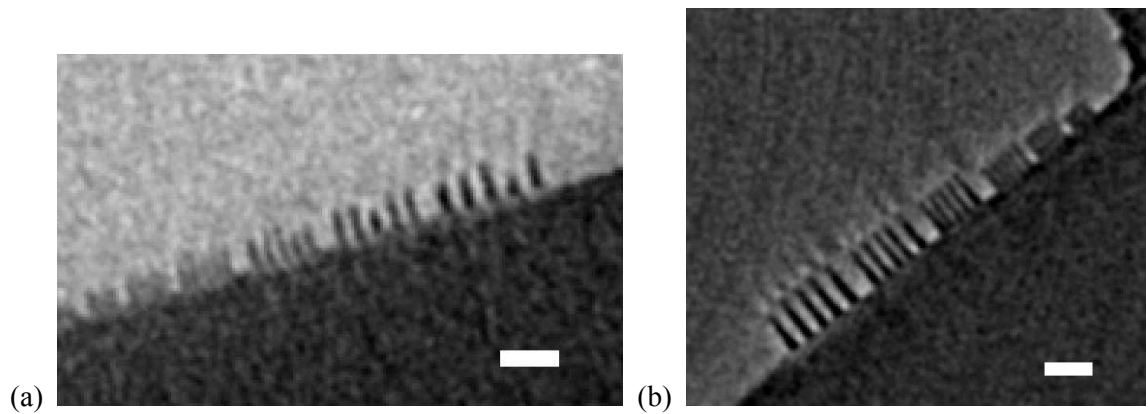

(a)   (b)

Figure 5. Microtomograms of square-wave patterns of aluminum (a) and diamond (b) test objects. Patterns with pitches of 2.0, 1.6, 1.2, 1.0, 0.8 and 0.6 μm were prepared, though only the 2.0, 1.6, 1.2 and 1.0 μm pitch patterns were resolved. LACs are shown in gray scale from -10 cm$^{-1}$ (black) to 50 cm$^{-1}$ (white). Scale bars: 5 μm.

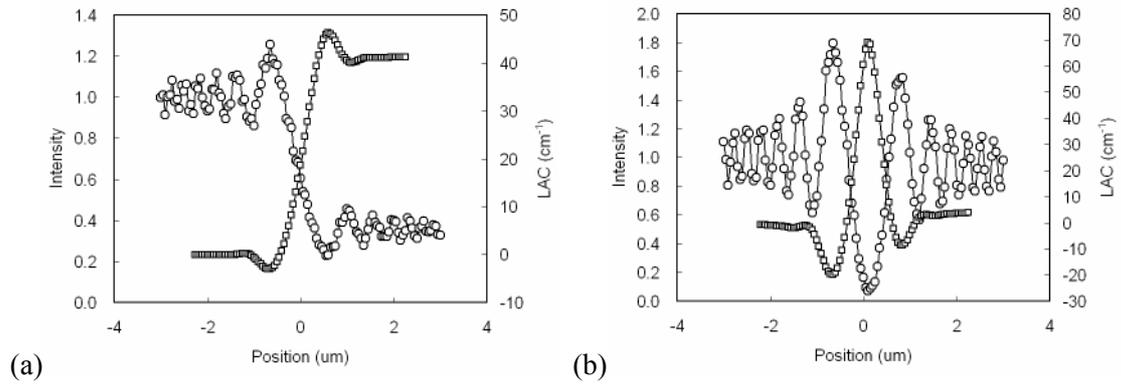

Figure A1. Kirchhoff integrals were calculated for a 250 μm edge of aluminum (a) and 150 μm edge of diamond (b). Simulated x-ray intensities are plotted with open circles. Step responses deduced from convolutions of the simulated intensities and observation LSF are plotted with open squares.